\documentclass[graybox]{svmult}
\usepackage[T1]{fontenc}
\usepackage[latin1]{inputenc}
\usepackage{graphicx}

\makeatletter

\begin{document}  

\title{Noise induced dynamics in adaptive networks with applications to
epidemiology}
\titlerunning{Noise induced dynamics in adaptive networks}

\author{Leah B.~Shaw and Ira B.~Schwartz}

\institute{Leah B. Shaw 
Department of Applied Science, College of William and Mary,
Williamsburg, VA 23187
\and Ira B. Schwartz
 US Naval Research Laboratory, Code 6792, Nonlinear Systems
Dynamics Section, Plasma Physics Division, Washington, DC 20375}

\maketitle

\abstract{
Recent work in modeling the coupling between disease dynamics and
dynamic social network geometry has led to the examination of how human
interactions force a rewiring of connections in a population.
Rewiring of the network may be considered an adaptive response to
social 
forces due to disease spread, which in turn feeds back to the
disease dynamics. Such epidemic models, called adaptive
  networks, have 
led to new dynamical instabilities along with the
creation of  multiple attracting states. The co-existence of several
attractors is 
sensitive to internal and external fluctuations, which lead to
enhanced stochastic oscillatory outbreaks and disease extinction.  The aim of
this paper is to explore the bifurcations of adaptive network models
in the presence of fluctuations and to review some  of the new
fluctuation phenomena induced in adaptive networks.}


\section{Introduction}

In recent years, researchers have used a network approach in studying many
systems, from networks of social contacts to the US power grid to the world
wide web \cite{Barabasi99}, and a wide variety of mathematical tools have been developed to
analyze static networks \cite{Newman03,CostaRTB07}. However, many natural systems are more complex than static network models.
Both the properties of individuals (e.g., neurons, humans) and the
connections between them change over time. Examples of networks
where the links evolve dynamically occur in simple two state models ,
such as two player game theory \cite{PachecoTN06,SkyrmsP00} and
opinion dynamics \cite{ZanetteG06,EhrhardtMV06}, as we will describe below.

Static network models fail to capture
systems in which dynamical properties are important.   A new class of models,
adaptive networks, has been 
introduced recently to address more fully the complexity of many physical
systems \cite{Gross07}.  In an adaptive network, the network geometry changes
dynamically in response to the node characteristics, and these changes in
geometry then alter the subsequent dynamics of the nodes. 

Many studies of adaptive networks have focused on steady state behavior, and
rich new phenomena have been discovered in that context.  However, the key
aspect of an adaptive network is the interplay of node dynamics and network
topology, which generally means that the nodes and links are evolving in time
even if a steady state is reached.  (Exceptions are cases where the network is
evolved to a frozen state, as in, for example, \cite{Gil06,Benczik07}.)  In
this chapter, we consider the role of fluctuations in adaptive network
models. 

One property that frequently occurs in adaptive networks is self organized
criticality, a topic that is discussed in more detail in the chapters of Rohlf and Bornholdt and of Caldarelli and Garlaschelli.  Because each node in the network receives dynamical information
that depends on the connectivity of the entire network, this can provide
global information to individual nodes and cause the system to organize
itself.  As a result, criticality and scale free behavior are often observed.
In the first adaptive network model, that of Christensen \textit{et al.}, adding and removing links to match a node's degree to the local average connectivity led to the network self organizing to a critical average connectivity, with a power law distribution of cluster sizes \cite{Christensen98}.  In a model by Bornholdt and R\"{o}hl  motivated by neural networks, the
network again organized to a critical average connectivity, balancing the addition
of new links when nodes are correlated and the removal of links when nodes are
uncorrelated \cite{Bornholdt03}.  Fan and Chen studied a growing network of
chaotic maps, in which new nodes were linked to the most active previous
nodes, and obtained scale free degree distributions \cite{Fan04}.  Zhou and
Kurths also obtained scale free distributions of connection weights for a
network of chaotic oscillators in which the weights were evolved to increase
synchrony \cite{Zhou06}.  Extensions to the latter two models are described in the chapter by Chen and Kurths.

None of
the above-mentioned studies looked at fluctuations, but scale free effects on fluctuations
have been observed previously.  Bornholdt and Sneppen studied an evolving
Boolean network, representing a genetic network, in which neutral mutations of
the couplings, those that do not affect the network attractor, accumulate over time
\cite{Bornholdt98}.  The average connectivity of the network was monitored,
and long periods of stasis in connectivity were interrupted by bursts of
connectivity change.  The stasis times followed a scale free distribution.  In
a model for an evolving network of chemical species, in which species that do
not multiply as quickly are replaced by new random species, Jain and Krishna
observed similar fluctuations in the number of populated species
\cite{Jain01}.  In the long time limit, all species are usually populated, but
this value is punctuated by drop-out times due to the disruption of
autocatalytic sets.  Jain and Krishna did not do a statistical analysis of
this effect, but their time series are reminiscent of punctuated equilibria. 

In evolutionary game theory models with perturbations, scale free
distributions have been observed for the sizes of avalanches (number of nodes
involved) as the system moves between stationary states (e.g.,
\cite{Ebel02,Scholz07}).  A model by Holme and Ghoshal for nodes that rewire
the network to maximize their social influence and change their rewiring
strategies adaptively also displayed avalanches in strategy changes, but
statistics were not collected on the scaling of the avalanche size
\cite{Holme06}.  The degrees and cluster sizes in this model also fluctuated
significantly.  This model is discussed in detail in the chapter by Holme and Ghoshal.

Another effect that is observed in adaptive network models for opinion
formation in social networks is the existence of metastable states.  Ehrhardt
and Marsili studied a model in which new links were generated preferentially
between nodes with a similar ``opinion'' or property, and the opinions were
influenced by neighbor nodes \cite{Ehrhardt06}.  In the limit where the nodes
were at zero temperature (but links were added and removed stochastically),
the system eventually approached one large connected component with uniform
opinion, but it spent time in metastable states with multiple components, each
with different opinion.  The lifetimes of the metastable states could be
understood analytically through an exact solution and stability analysis of
the zero temperature case. 

Holme and Newman developed a model for opinion dynamics in which a parameter
governed the relative frequencies of rewiring to new neighbors with identical
opinions versus convincing one's neighbors to share one's opinion
\cite{HolmeN06}.  The system evolved to a frozen state containing one or more
communities, where the number of communities depended on whether the parameter
favored convincing or rewiring.  The parameter controlled a continuous phase
transition, and at the critical value the system exhibited a power law
distribution in community sizes and large fluctuations in the time required to
converge to the final state.  The nature of the phase transition was further
explained analytically by Vazquez et al.~for a simpler model with only two
opinion states \cite{Vazquez08}.  A variety of models of opinion formation are described in detail in the chapter by Do and Gross.

Other adaptive network models have considered synchrony of a network of
coupled oscillators while adjusting the network connections adaptively \cite{Ito02,Ito03,Gong04,Gleiser06,Zhou06}.  Gong and
Leeuwen studied networks of coupled chaotic maps and added connections between
correlated oscillators \cite{Gong04}.  They found that the system formed a
small world network with intermittent switching in the number of coherent
clusters.  Ito and Kaneko studied weighted networks of coupled maps and also
strengthened the coupling between correlated oscillators
\cite{Ito02,Ito03}. Here they considered in detail the role of the feedback mechanism between node and network dynamics.  The model is also discussed in the chapter by Ito and Kaneko.  They computed
average weight 
matrices to determine whether there were stable structures in the network and
found several phases, including a phase that was desynchronized but had a
temporarily stable network structure, and a desynchronized phase with a
disordered, rapidly changing network structure \cite{Ito03}.  In the
disordered network structure, the degree of individual nodes changed almost
randomly.  However, in the networks with temporarily stable structure, the
nodes separated into two groups, controllers with high outdegree and others
with low outdegree.  The group in which a given node resided remained
relatively stable over time.  The node dynamics in the desynchronized phase
was characterized by hopping between two groups whose dynamics were a half
cycle out of phase from each other.  In a feedback loop between the network
geometry and node dynamics, nodes that hopped more slowly between groups
tended to accumulate more connections, which led to further slowing of the
hopping rate.  This feedback loop was responsible for the splitting of the
nodes into high outdegree and low outdegree groups. 

Effects such as self organized criticality, metastable states, and
fluctuations in synchrony have been observed in adaptive networks.  Thus far,
these effects have mainly been quantified through simple network metrics such
as time series of the average connectivity or clustering coefficient, or the
average node properties may be tracked over time as will be discussed later in
this chapter.  When the network forms distinct clusters, tracking the number of clusters is also an option. However, higher order network structures are often difficult to track
in a time-varying network, and it is not yet clear what are the key network
properties to measure for an adaptive network.  Also, in many cases the
effects that have been observed have not yet been explained analytically.
Further study of the fluctuations in adaptive networks is needed. 

In this chapter, we focus on fluctuations in a model for an epidemic spreading
on an adaptive network.  Epidemics have been briefly mentioned in the previous chapter.  Some of the results in the present chapter have been published elsewhere
\cite{Shaw08}. The layout of the paper is as follows: We describe the model and summarize key aspects of its
bifurcation structure in Sections \ref{sec:model} and \ref{sec:bifurcation}.
Many properties can be predicted from a much lower dimensional mean field
model.  In Section \ref{sec:fluctuations}, we focus particularly on
fluctuations in the number of infection cases in the system, which is a
physically important quantity.  We present some additional results for phase
relationships between node and link variables and for scaling of the epidemic
lifetime in Sections \ref{sec:delayedoutbreaks} and \ref{sec:lifetimes}.  The
dynamic network structure is more difficult to capture, but in Section
\ref{sec:network} we discuss fluctuations in the degree of individual nodes in
the system. 

\section{Model}

\label{sec:model}

Gross \textit{et al}.~have introduced a susceptible-infected-susceptible
(SIS) model on an adaptive network \cite{GrossDB06}, and Zanette and Gusm\'{a}n have also studied an SIS model on an adaptive network \cite{Zanette07}.  We have
extended this work to a susceptible-infected-recovered-susceptible
(SIRS) model \cite{Shaw08}.  Although we have not chosen
parameters corresponding to a particular real disease, tuning the
average time a node spends in the recovered class allows us to
adjust the average number of infections at the endemic steady
state. Although some diseases in the past, such as plague, have eliminated as
much as 50 per cent of a worldwide population,  many  infectious viral diseases, such as measles, mumps, and rubella, infect only 10\% or less of a population at a
given time \cite{Anderson91}, depending on epidemiological and social factors.  Noise effects are expected to be especially prominent when the
infection occurs at low levels. However, in this particular study,
we restrict our attention to cases where the minimum endemic steady states
are on the order of 10-40\% of the population.

Our model is constructed as an extension to that of Gross et
al.~\cite{GrossDB06}  but includes the addition of a recovered class.
The rate for a susceptible node to become infected is $p
N_{I,\textrm{nbr}}$, where $N_{I,\textrm{nbr}}$ is the number of
infected neighbors the node has.  The recovery rate for an
infected node is $r$.  We fix $r=0.002$ throughout this chapter.
A recovered node becomes susceptible again with rate $q$, the
resusceptibility rate. 

Since the mean time spent in the recovered state is $1/q$, there is a
natural limiting case for the SIRS model. Using the recovery rate, $r$, as a
natural time scale, as the ratio $q/r$ becomes sufficiently large,  nodes spend less
time in the recovered state. In the limit  $q/r \rightarrow \infty$,   the
model thus  approaches the SIS model. The study of the stochastic
dynamics of the SIRS model may then be examined with respect to changes in
$q$. 

Rewiring of the network occurs as the epidemic spreads. If a link
connects an infected node to a non-infected node, the link is
rewired with rate $w$.  The connection to the infected node is
broken, and the original non-infected node is now connected to
another non-infected node which is randomly selected out of all
candidates in the network (excluding self links and multiple links
between nodes).  This rewiring rule is that of Gross et al.~\cite{GrossDB06}
(we treat the recovered nodes in the same manner as susceptibles for rewiring
purposes), in contrast to the rewiring scheme of Zanette and Gusm\'{a}n
\cite{Zanette07}, which allows susceptible nodes to connect to infectives.  

We performed Monte Carlo simulations of this model for a system
with $N$ nodes and $K$ links, where $K/N$ was fixed at 10. Details
of the simulation procedure can be found in \cite{Shaw08}.  Random
sequential updating was used, and each node and eligible link had
an opportunity to transition on average once per Monte Carlo step
(MCS).

As in \cite{GrossDB06}, we developed a corresponding mean field
model using a moment closure approximation to track the dynamics
of nodes and links. $P_A$ denotes the probability for a node to be
in state $A$, and $P_{AB}$ denotes the probability for a link to
connect a node in state $A$ to a node in state $B$. For higher
order correlations, we assume $P_{ABC} \approx P_{AB} P_{BC} /
P_B$. The time evolution of the node states is described by:
\begin{eqnarray}
\label{eq:mf_node}
\dot{P}_S &=& q P_R -p \textstyle{\frac{K}{N}} P_{SI} \\
\dot{P}_I &=& p \textstyle{\frac{K}{N}} P_{SI}-r P_I \\
\dot{P}_R &=& r P_I -q P_R
\end{eqnarray}
The time evolution of the links is described by:
\begin{eqnarray}
\dot{P}_{SS} &=& q P_{SR} +w \textstyle{\frac{P_S}{P_S+P_R}} P_{SI} -2p \textstyle{\frac{K}{N}} \frac{P_{SS}P_{SI}}{P_S} \\
\dot{P}_{SI} &=& 2p\textstyle{\frac{K}{N}} \frac{P_{SS}P_{SI}}{P_S} +q P_{IR} -r P_{SI} -w P_{SI} \nonumber \\
&& -p \left( P_{SI} + \textstyle{\frac{K}{N}} \frac{ P_{SI}^2}{P_S} \right) \\
\dot{P}_{II} &=& p \left( P_{SI} + \textstyle{\frac{K}{N}} \frac{ P_{SI}^2}{P_S} \right) - 2r P_{II} \\
\dot{P}_{SR} &=& r P_{SI}+w \textstyle{\frac{P_R}{P_S+P_R}} P_{SI} +2q P_{RR}-qP_{SR} \nonumber \\
&& -p \textstyle{\frac{K}{N}} \frac{P_{SI} P_{SR}}{P_S} +w \frac{P_S}{P_S+P_R} P_{IR} \\
\dot{P}_{IR} &=& 2r P_{II} +p \textstyle{\frac{K}{N}} \frac{P_{SI} P_{SR}}{P_S} -q P_{IR} -r P_{IR} \nonumber \\
&& -w P_{IR} \\
\dot{P}_{RR} &=& r P_{IR} -2q P_{RR} +w
\textstyle{\frac{P_R}{P_S+P_R}} P_{IR} \label{eq:mf_links}
\end{eqnarray}
We integrated the mean field equations numerically and tracked
their steady states using a continuation package \cite{auto}. We
have also considered a stochastic mean field system with internal
fluctuations, modeled by multiplicative noise, or with external
fluctuations, modeled by additive noise. The stochastic mean field
system was studied using a fourth order Runge-Kutta solver.

\section{Bifurcation structure}

\label{sec:bifurcation}

In \cite{Shaw08}, we considered the bifurcation structure of the mean field
for $q$ small and mapped the regions of stability for infectives as a
functions of the parameters $w$ and $p$. We discovered and reported that there
were regions of $q$ in which different bifurcation scenarios existed, as well
as regions of bistability. In the first case, the value of $q=0.0064$, the
disease free equilibrium became unstable as the infection rate $p$ was
  increased, and through a transcritical bifurcation was connected to
the unstable branch of endemic states. A stable branch of endemic states was then
connected to the unstable branch via a saddle-node bifurcation. However, in
the second case, that of $q=0.0016$, there is no saddle-node
  bifurcation. Instead there exists a saddle-saddle connection. In
    this case, the unstable endemic state emanating from the disease free
    state has a one-dimensional unstable manifold. This branch is connected via
  a turning point to an unstable endemic branch having a  two-dimensional  unstable manifold,
  which in turn becomes stable through a Hopf bifurcation.
 
\begin{figure}[tbp]
\begin{center}
\includegraphics[width=3.0in,keepaspectratio]{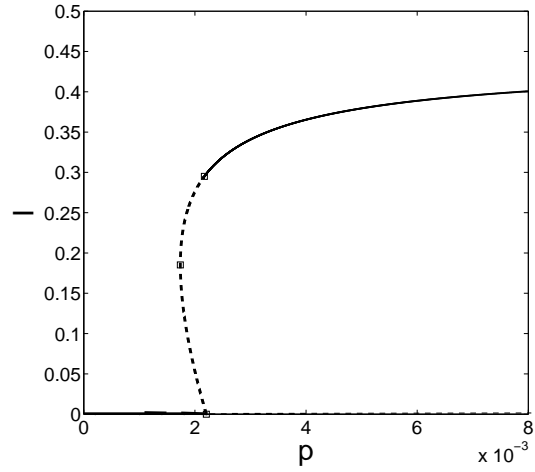}
\end{center}
\caption{  A
 a bifurcation diagram of the infective fraction  as a function of $p$, with $w =
 0.04$, $r=0.002$, $ q=0.0016$. The squares denote the saddle-saddle point and transcritical
point. Dashed lines are unstable branches. As $p$ is decreased, the endemic state loses stability in a Hopf bifurcation.}
\label{fig:BDparamq0016}
\end{figure}

 The bifurcation diagram for this case is
reproduced here for clarity of discussion in Fig.~\ref{fig:BDparamq0016}.
For the saddle-saddle case, the lower (upper) branch has a one (two) dimensional unstable manifold. The upper branch then
undergoes a reverse Hopf bifurcation. The connecting branch of periodic orbits
(not shown) is unstable and sub-critical. These orbits have very long
periods and large swings in amplitudes of infectives.

\begin{figure}[tbp]
\begin{center}
\includegraphics[width=3.0in,keepaspectratio]{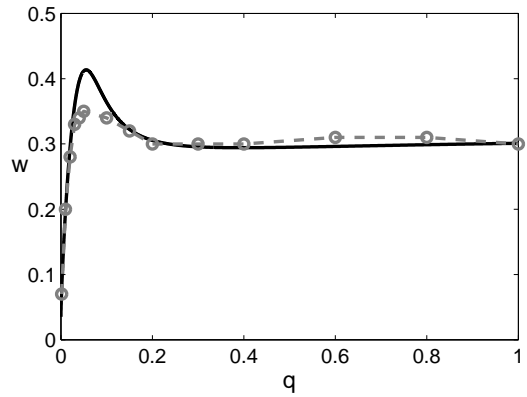}
\end{center}
\caption{  A two parameter
  diagram of the Hopf bifurcation points as a function of  $q$ and
  $w$.  Solid curve:  mean field model; points and dashed curve: full system.  Parameters used are $r=0.002,p=0.004.$ 
}
\label{fig:HB_wq}
\end{figure}

Recalling that  $q$ controls the resuscepibility rate  and $w$ the rewiring
rate, we examine the structure of the
bifurcation onset of attracting endemic states while holding the other parameters
fixed.  The onset of Hopf
bifurcation points in two parameters was computed, and is  shown in Fig.~\ref{fig:HB_wq}, for both the mean field model and the full system. Bifurcation points in the full system were estimated as the largest $w$ value for which a single run started near the probable endemic steady state remained near that state for $10^5$ MCS without dying out.  The
region below the curve contains stable endemic branches, while the region
above  contains unstable endemic states and/or stable disease free
equilibria. The same trends are observed in both the mean field and the full system. Notice that for $q$
greater than approximately $0.3$, the value  of $w$ for the Hopf bifurcation does not change
 much. In addition, the infective fraction is also approximately independent of $q$
 for $q$ sufficiently large,  signifying that the model is approaching the SIS
 model. One would typically expect that as the resusceptibility rate $q$
 increases, the number of nodes that are in the recovered state and thus
 protected from infection will decrease, and the infection will spread more
 easily.  Therefore, a faster rewiring rate (larger $w$) will be needed to
 suppress the infection.  Indeed, this is the trend observed  in
 Fig.~\ref{fig:HB_wq} for small and large $q$.  However, the $w$ value for the
 bifurcation decreases with increasing $q$ between about 0.1 and 0.2.  This
 nonmonotonic shift in the bifurcation point in Fig.~\ref{fig:HB_wq} is a
 nonlinear effect which has not yet been explained. 

In later
 sections we will explore
 the fluctuations of the SIRS model for values of $q$ between the values
 0 and 1.  Effects on the fluctuations of the nonmonotonic bifurcation curve in Fig.~\ref{fig:HB_wq} have not been observed.

We remark that a
direct comparison in \cite{Shaw08} of the infective fraction between the  mean field model and Monte Carlo simulation of the full system showed excellent agreement along the attracting
branches. The discrepancies occurred near the bifurcation point where the endemic state loses stability, partly because the
actual location of the instability in the full stochastic system  was difficult to detect accurately in Monte Carlo simulations and partly due to inaccuracies in the mean field approximation.  However, the scaling results near the bifurcation, which we present in the next section, are generally consistent between the mean field model and the full system.

\section{Effect of recovered class on fluctuations}

\label{sec:fluctuations}

We have observed that the amplitude of fluctuations in the number
of infectives is generally larger in the SIRS model than the SIS
model.  In the SIS system, links between two infectives are not
broken because rewiring operates only on SI links.  Thus when an
infective becomes susceptible again, the newly formed susceptible
may be connected to other infectives that it retained as neighbors
while previously infected and can immediately become reinfected.
This situation tends to suppress fluctuations, because small
decreases in the total number of infectives correspond exactly to
increases in the number of susceptibles, and rapid reinfection of
the susceptibles can occur, preventing the number of infectives
from dropping significantly. In the SIRS model, on the other hand,
the recovered compartment introduces an effective time delay from
recovery to possible reinfection and allows infective levels to
fluctuate more.

\begin{figure*}[tbp]
\begin{center}
\includegraphics[width=5in,keepaspectratio]{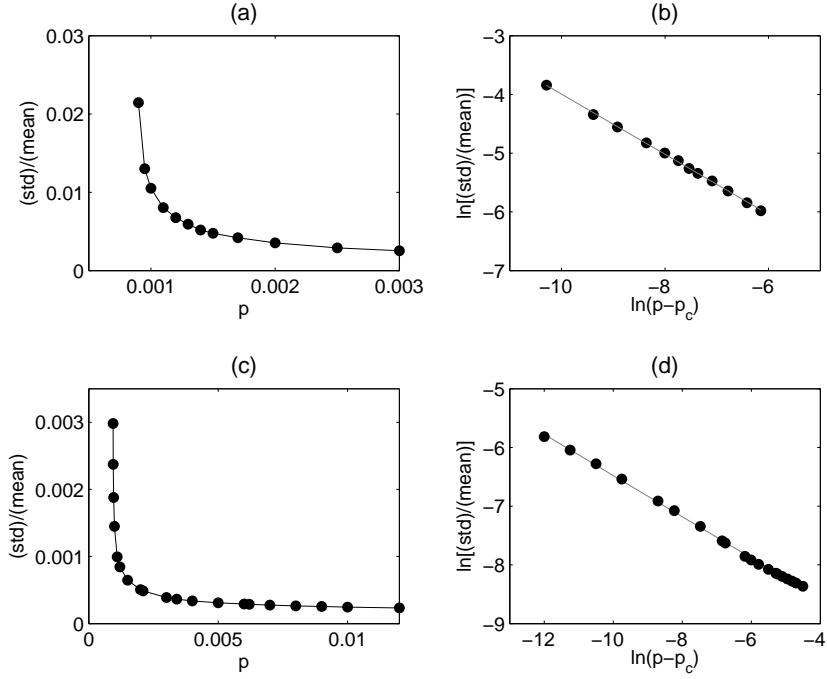}
\end{center}
\caption{Fluctuations in infectives (standard deviation divided by
mean) vs.~infection rate $p$ near the bifurcation point, from
Monte Carlo simulations:  (a) $q=0.0016$ (SIRS case), (c) $q=1$
(approximately SIS case).  Curves are to guide the eye.  Log-log
plots (data points with best fit lines) show power law scaling for
both $q=0.0016$ (b) and $q=1$ (d).  Other parameters:  $w=0.04,
r=0.002$.  Parts (a) and (b) are reprinted from \cite{Shaw08}.}
\label{fig:fluctvsp}
\end{figure*}

Figure \ref{fig:fluctvsp} compares the scaling of fluctuations
near the bifurcation point for two different values of the
resusceptibility rate $q$.  In the top panels $q=0.0016$, the rate
used in \cite{Shaw08}.  In the bottom panels $q=1$, effectively
approximating the SIS case, since individuals spend very little
time in the recovered class and much less than 1\% of the
population is in the recovered class at a given time. Fluctuations
in the infectives (measured as the standard deviation divided by
mean for long Monte Carlo simulations) are plotted as a function
of $p$, the infection rate, as $p$ is swept towards the
bifurcation point. Results were computed from $5 \times 10^5$ MCS
time series sampled every 10 MCS.  The magnitude of the
fluctuations is greater for the SIRS case (Fig.~\ref{fig:fluctvsp}a)
than for the SIS case (Fig.~\ref{fig:fluctvsp}c). Notice that the increase
  in fluctuations is almost an order of magnitude.

The
fluctuations exhibit power law scaling, shown in the log-log plots in
Fig.~\ref{fig:fluctvsp}b and \ref{fig:fluctvsp}d.  On the horizontal axis, we plot
$\ln (p-p_c)$, where $p_c$ is the critical point at which the endemic state
loses stability.  The bifurcation points are not known exactly, so we approximate  $p_c$ by the value that
produces the most linear plot in each case.  For the data of Figure \ref{fig:fluctvsp}, the scaling exponents are similar (-0.59 versus -0.51 for $q=0.0016$ and 1 respectively), so it is not clear that the resusceptibility rate has a significant effect on how the fluctuations scale with $p$.

\begin{figure}
\includegraphics[width=2.3in]{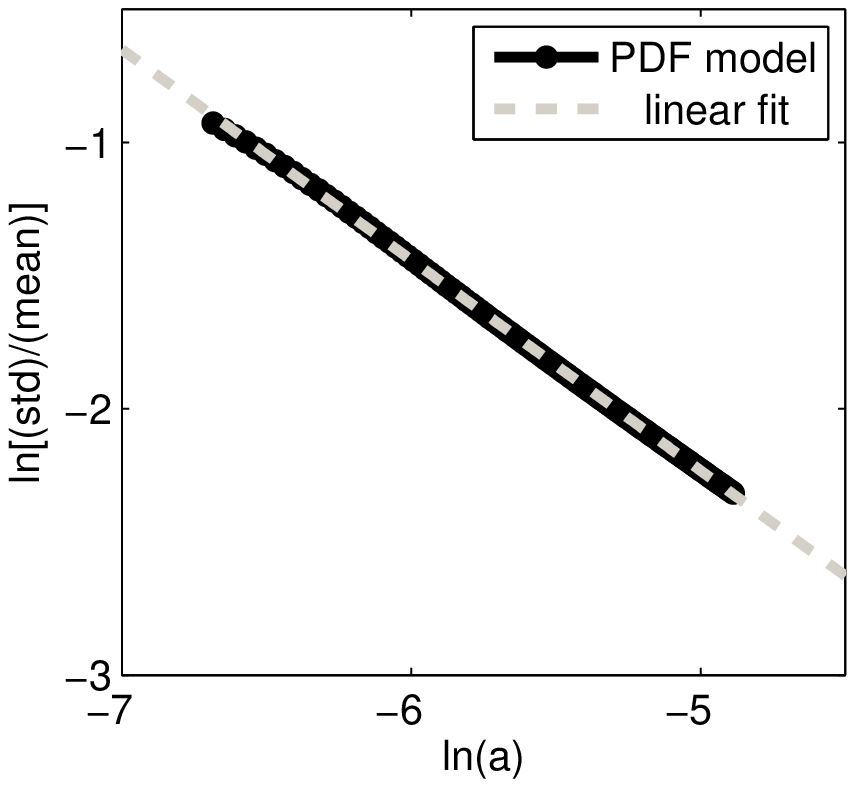}
\includegraphics[width=2.3in]{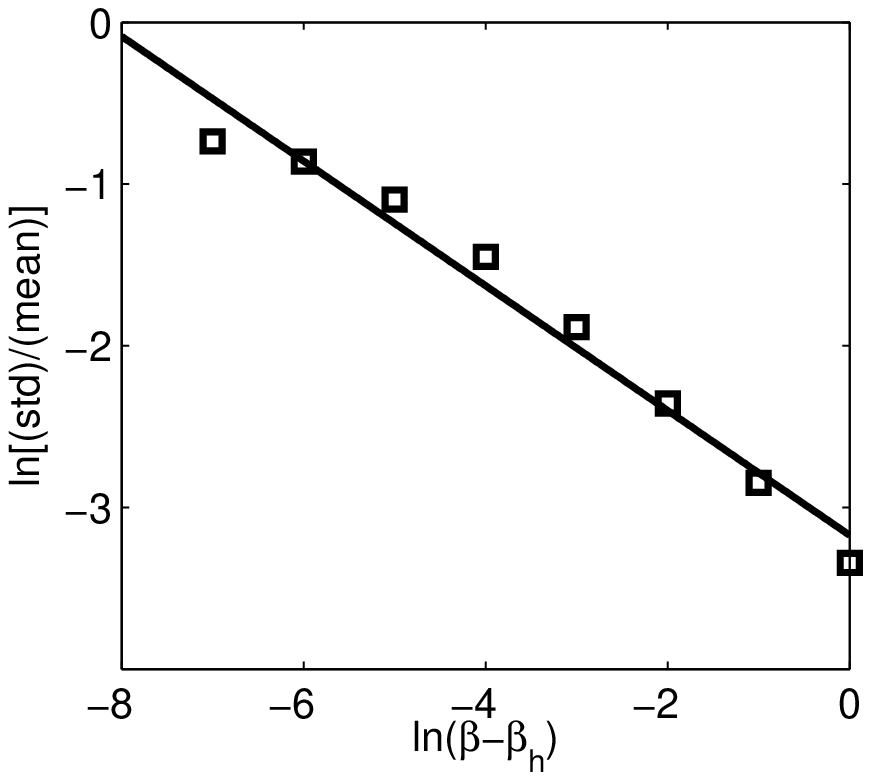}

\caption{\label{fig:SN_bifurcation_fluctuations} (a) Fluctuation size of a generic
saddle-node bifurcation as a
function of bifurcation parameter $a$ near the bifurcation point using the probability density function in
Eq.~\ref{Eq.SNPDF}. Noise amplitude is $\sigma=0.005$. (b) Fluctuation size of a generic
Hopf bifurcation as a
function of bifurcation parameter $\beta$ near the bifurcation point using the
PDF in Eq.~\ref{Eq:HBpdf} (squares). The line is a best linear fit. Noise
amplitude is $\sigma=0.05$. }

\end{figure}

The power law scaling of the fluctuations can be understood by
considering the scaling near a generic bifurcation point.  From
our mean field analysis, we expect the bifurcation point where the
endemic steady state loses stability to be either a saddle-node
bifurcation point or a Hopf point.  A generic saddle-node
bifurcation exhibits power law scaling of fluctuations near the
bifurcation point, as we show in Fig.~\ref{fig:SN_bifurcation_fluctuations} and explain in the discussion below. A Hopf bifurcation can also
appear locally to have power law scaling of fluctuations, although
the scaling may be over smaller range of parameters. For a given standard
deviation, the probability density function near a Hopf bifurcation is given by \cite{arnold}:
\begin{equation}
p_{hb}(\beta, r, \sigma, R)=N  {r}^{\,{\frac {\beta}{{\sigma}^{2}}}}{e^{- {\frac {R{r}^{2}}{{\sigma}^{2}}}}}{\sigma}^{-2} \left[ \Gamma  \left(  1+1/2\,{\frac {\beta}{{\sigma}^{2}}} \right)  \right] ^{-1}
\label{Eq:HBpdf}
\end{equation} 
where the state space variable is a radial coordinate $r$, $\beta$ is the distance from the Hopf bifurcation
  point, parameter $R=0.5$ is fixed,  $\Gamma$ is the gamma function, and $N$ is a normalization constant.  An example of the fluctuations  using  
Eq.~\ref{Eq:HBpdf}, where  we have computed the first and second
order moments to find  the ratio of the
standard deviation to the mean, is
shown in Fig.~\ref{fig:SN_bifurcation_fluctuations}b. The data for the  fluctuation size deviates from a power law
scaling, as exhibited in the figure. However, there is a monotonic relationship in
the fluctuations as measured by $\sigma / \mu$ as a function of the distance to
the Hopf bifurcation. Therefore, we expect the fluctuation characteristics of
the SIS and SIRS models to hold near bifurcation points regardless of whether the statistics are
measured with respect to a Hopf or saddle-node bifurcation.

Since the power law scaling is observed near either a saddle-node or Hopf
bifurcation point, it may be understood by considering the local dynamics. For
example, near the saddle-node, a center manifold reduction would reduce the
study of the vector field to a system with a one-dimensional unstable manifold. Therefore, 
the power law scaling of the fluctuations can be motivated by considering the following simple stochastic differential equation
\begin{equation}
dx_{t}=(a-x_{t}^{2})dt
+\sigma*dW_{t}
\label{Eq.SNSDE}
\end{equation}
for a one dimensional saddle-node bifurcation, where $a$ is the bifurcation parameter, $dW/dt$ is a white noise
term, and $dW$ is a Brownian increment.  In general, noise can cause a shift in the location of the saddle-node bifurcation, so we assume that the noise is sufficiently small that the location is fixed.

By assuming  we are always
near the attracting branch of the saddle-node or Hopf bifurcation,  we are in a near
equilibrium setting driven by noise. Such an assumption allows us to examine the
stationary probability density function (PDF) of the stochastic
dynamics by employing  the Fokker-Planck equation near steady
state. For the stochastic differential equation,
Eq.~\ref{Eq.SNSDE}, the PDF is well known \cite{Horsthemke83} and
is given by
\begin{equation}
p(a,x,\sigma)=Ne^{2(a x-x^{3}/3)/\sigma^{2}}.
\label{Eq.SNPDF}
\end{equation}
Here $N$ is a normalization constant. From Eq.~\ref{Eq.SNPDF}, we compute the first and second
order moments to find  the ratio of the
standard deviation to the mean.  We examine the fluctuations in the neighborhood
of $a=0$, which is the location of the saddle-node point. The results display power law
scaling, as depicted in
Fig. \ref{fig:SN_bifurcation_fluctuations}a. 

To further examine the
differences in fluctuations between the SIS and SIRS adaptive
network models, we perform the following experiment. As discussed
above, we can examine the fluctuation sizes as a function of resusceptibility rate $q$
to see how the fluctuation sizes compare between the two model
classes. The other variable which controls the recovered, as well
as the infected, populations is the rewiring rate, $w$.  It has
a significant effect on the fluctuations, since the degree of
infectives is dramatically reduced by the rewiring.  We examine
the interplay between $q$ and $w$ and their effect on fluctuation
sizes.  Here we turn to a stochastic version of the mean field
model.  We have shown previously that the scaling behavior of the
mean field model is typically similar to that of the full network
system \cite{Shaw08}.

We use additive noise to model fluctuations near the endemic
equilibrium state.  Details may be found in \cite{Shaw08}. The
stochastic mean field model has the following form:
\begin{equation}
\bf{X}' = \bf{F(X)} + \epsilon \bf{\eta} (t),
\end{equation}
where $\bf{F(X)}$ is the mean field system in
Eqs.~\ref{eq:mf_node}-\ref{eq:mf_links}, $\eta(\bf{t})$ is a noise term with    $\langle \bf{\eta} (t)
\bf{\eta} (t') \rangle = \delta (t-t')$, and $\epsilon$ is the
noise amplitude. Fluctuations are computed by averaging the
standard deviation over mean results for time series starting from  10 random
initial conditions near steady state. The runs were computed for $5 \times 10^7$
steps using a step length of 0.001, and transients were removed after $10^6$
steps.

\begin{figure}
\includegraphics[width=2.3in]{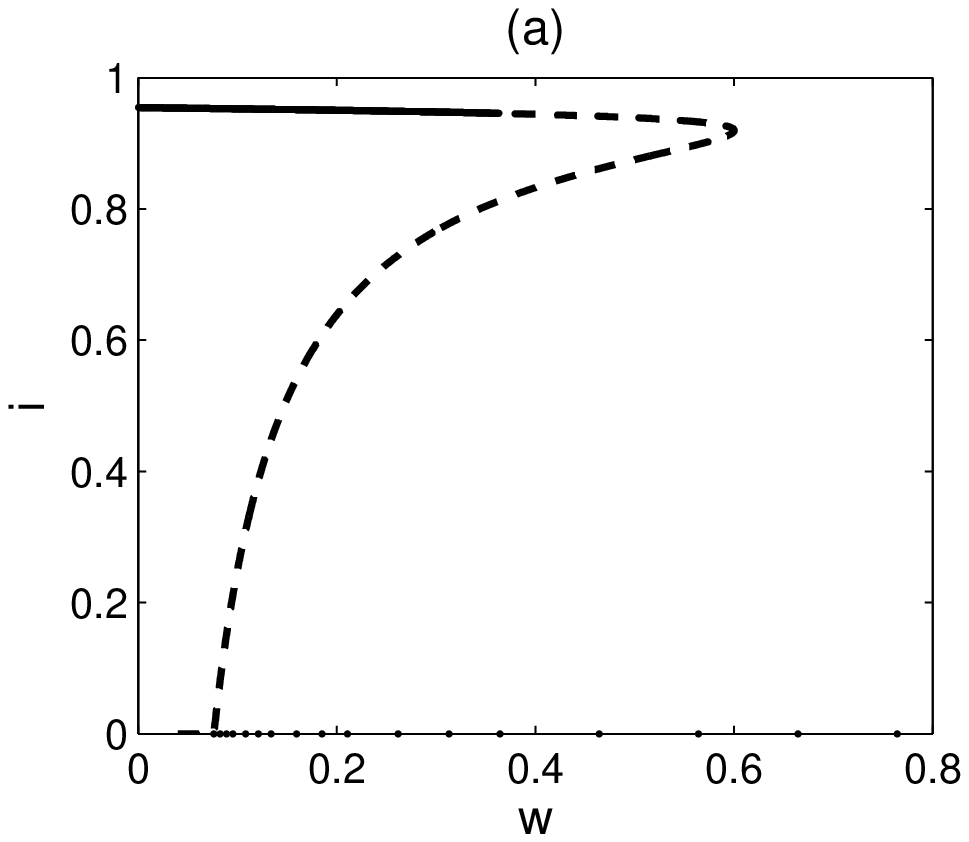}
\includegraphics[width=2.3in]{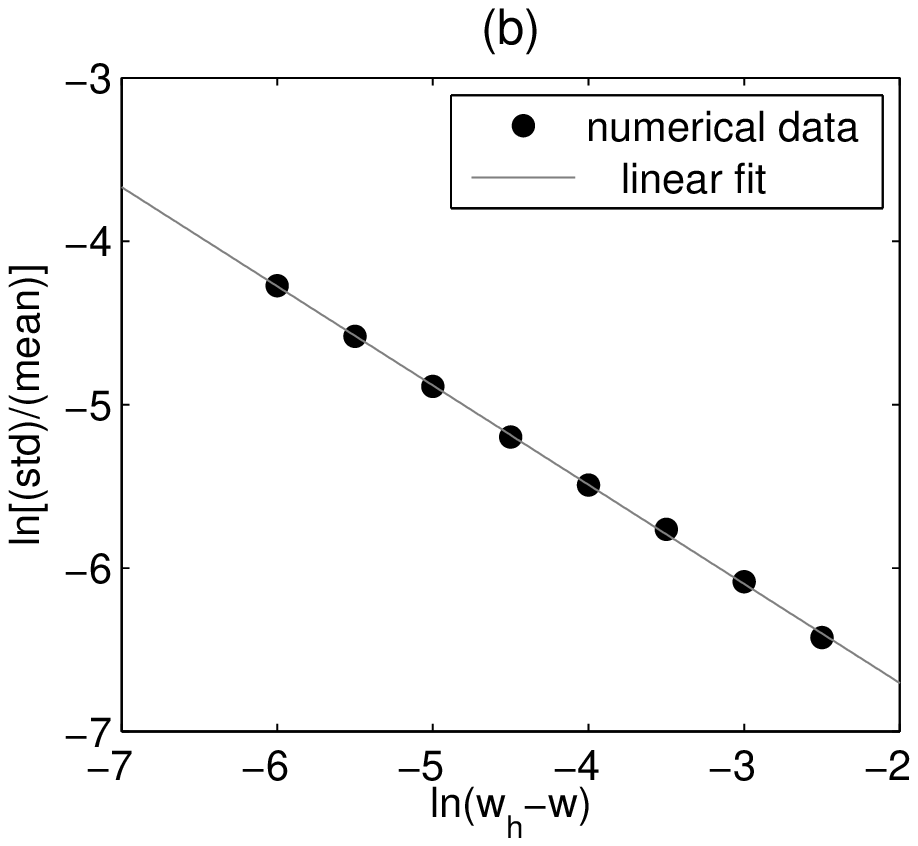}

\caption{(a) A bifurcation plot of the mean field model without noise.
Plotted is the fraction infected as a function of $w$. Parameters are $r=0.002, p=
0.004,  q=0.1$. (b) Plot of the fluctuation sizes as a function of rewiring
rate, $w$.  The fixed parameters used are $r=0.002,\, p=0.004,\:
K/N=10,\: q=0.01, \epsilon = 0.0001$.} 

\label{fig:BP_log_flucts}

\end{figure}

A typical example of the fluctuations as $w$ is varied is shown in
Fig.~\ref{fig:BP_log_flucts}b. Similar linear log-log behavior is
observed in other stochastic simulations for other values of $q$.
In Fig.~\ref{fig:BP_log_flucts}b, $w_{h}$ denotes the location of the Hopf bifurcation
branch. The Hopf bifurcation occurs for all values of $q$
considered here. A typical bifurcation plot is shown in Fig.
\ref{fig:BP_log_flucts}a for $q=0.1$. Attracting states are solid curves,
while unstable states are dashed and dotted curves. The Hopf bifurcation point is on
the upper branch separating the stable and unstable steady states.

\begin{figure}
\begin{center}
\includegraphics[width=2.5in]{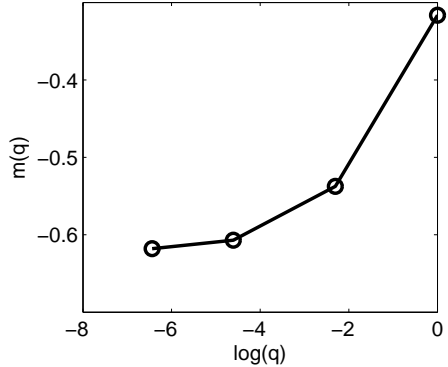}
\end{center}

\caption{A plot of the slope $m(q)$ as a function of $q.$  See text for details.
Parameters used are $r=0.002, p=0.004, \epsilon = 0.0001$.} \label{fig:compiled_slopes}

\end{figure}

Because of the power law scaling of the fluctuations, as in Figure
\ref{fig:BP_log_flucts}b, we expect a functional relationship of
the form
\begin{equation}
\sigma(q)/\mu(q) \propto \left[w_{h}(q)-w\right]^{m(q)}\label{eq:HBscaling}
\end{equation}
where $m(q)$ is the average slope of the log-log plots. We can now
examine how the average rate of change of fluctuations varies as
a function of $q.$ The results are shown in Fig.~\ref{fig:compiled_slopes}.  At smaller $q$ values, the fluctuations increase more
quickly with $w$ than they do in the large $q$ limit. Therefore, the fluctuations are
  more sensitive with respect to $w$ in the SIRS model than in the SIS model. 

  We attempted to confirm this mean field result for the slopes using
  the full model, but in the case of the full model, the exact locations of
  bifurcation points are unknown.  It is difficult to estimate where the
  endemic state loses stability from time series because one cannot always
  distinguish a metastable state from a stable state in the presence of
  fluctuations.  We can approximate the bifurcation point by the value that
  gives the most linear plot (largest $R$ value) for fluctuations vs.~the
  bifurcation parameter as in Figure \ref{fig:fluctvsp}, but this approach is
  unreliable if the scaling deviates from a power law, as occurs in Figure
  \ref{fig:SN_bifurcation_fluctuations}b for a generic Hopf bifurcation.  The
  best fit slope depends sensitively on the estimate for the bifurcation
  point, so we were not able to obtain robust results for the full system.

\section{Delayed outbreaks}

\label{sec:delayedoutbreaks}

\begin{figure}[tbp]
\begin{center}
\includegraphics[width=3in,keepaspectratio]{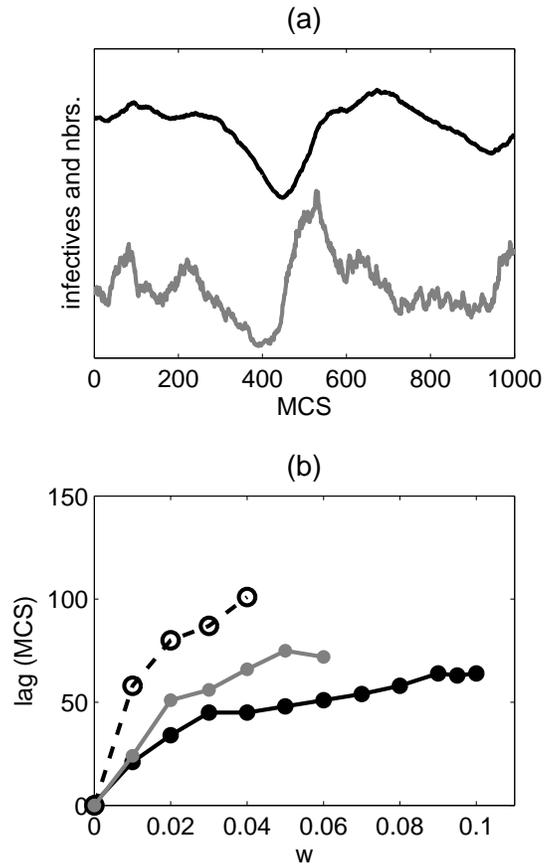}
\end{center}

\caption{Delayed outbreaks due to rewiring.  (a) Monte Carlo time series.  Black: infectives; gray:
neighbors of infectives.  Curves are scaled in arbitrary units for comparison
of peak times.  $p=0.0065, w=0.09, q=0.0016, r=0.002$. Reprinted from \cite{Shaw08}. (b) Time in MCS by
which infectives lag behind infective neighbors vs.~rewiring rate.  Solid
black:  $q=0.0016$; gray:  $q=0.001$; dashed black: $q=0.0005$.  Other parameters:  $p=0.0065,
r=0.002$.
}
\label{fig:lags}
\end{figure}

We have also considered phase relationships between the
fluctuating node and link variables.  At each time point in our
simulations, we tracked the number of infected nodes as well as
the number of non-infected neighbors of infected nodes (which
corresponds to the number of SI and IR links).  The rewiring
causes the fluctuations in the number of infectives to lag behind
fluctuations in the number of infective neighbors, as shown in
Figure \ref{fig:lags}a.  This effect was observed both in the mean
field model (data not shown, see \cite{Shaw08}) and in Monte Carlo
simulations of the full system.

We studied the dependence of this phase lag on both the rewiring
rate $w$ and the resusceptibility rate $q$ when the system was
fluctuating around the endemic steady state.   Monte Carlo
simulations were sampled every 1 MCS for $3 \times 10^4$ MCS after
discarding transients.  We computed cross correlations between the
infectives and the infective neighbors for varying phase shifts
between the two time series and identified the lag maximizing the
cross correlation.  Figure \ref{fig:lags}b shows results for three
different $q$ values.  (Note:  Curves for smaller $q$ terminate at
lower $w$ values because the endemic steady state becomes
unstable, as in Figure \ref{fig:HB_wq}.)  In each case, the lag increases with increasing
rewiring rate.  This effect does not have a simple explanation,
but since it is also observed in mean field simulations, it
depends on node and link dynamics primarily, rather than higher
order geometries.

Further, the lag time increases as the resusceptibility rate $q$ decreases.  This occurs because the recovered class introduces an effective delay in the system.  When a node becomes at risk because its neighbor is infected, it cannot itself become infected until it is susceptible.  As $q$ is lowered, the fraction of infective neighbors that are recovered and have to wait to become susceptible again increases, and the average wait time also increases, so it is expected that the infective fluctuations will lag further behind.

It should be noted that when $q=1$ and the system approximates the SIS model, the number of infectives and non-infected neighbors of infectives (i.e., SI links) are poorly correlated for any shift between time series.  Therefore, the lags discussed here are not observed in the SIS model.

\section{Lifetime of the endemic steady state}

\label{sec:lifetimes}

\begin{figure}[tbp]
\begin{center}
\includegraphics[width=3in,keepaspectratio]{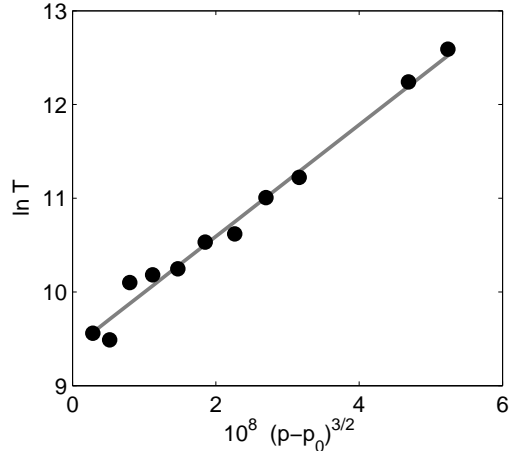}
\end{center}
\caption{Dependence of endemic state average lifetimes $T$ on
infection rate $p$. Points:  Monte Carlo simulations; line:  best
fit line.  $q=0.0016$, $r=0.002$, $w=0.04$, $N=4 \times 10^4$, $K=4\times 10^5$.  (Reprinted from
\cite{Shaw08}.) } \label{fig:lifetimes}
\end{figure}

Another effect we consider, which depends on fluctuations in the system, is the
lifetime of the endemic steady state.  Because the system is
stochastic and the disease free state is absorbing, the disease will die out in the infinite time limit for any set of parameters.
For a generic saddle-node bifurcation in one dimension, the scaling of the lifetime is expected to obey
\begin{equation}
\ln {T} \propto (p-p_0)^{3/2},
\end{equation}
where $T$ is the mean dwell time or lifetime of the steady state, $p$ is the bifurcation parameter, and $p_0$ is the location of the
bifurcation point \cite{Dykman1980,Graham1987a}.
Using the computational methods in \cite{Shaw08}, we show preliminary results for the dependence of the lifetime on the infection rate $p$ in Figure \ref{fig:lifetimes}.  The bifurcation point $p_0$ was estimated by the value that gave the most linear plot for $\ln T$ vs.~$ (p-p_0)^{3/2}$.  The scaling
results appear consistent with expectations, but further study is needed.  In contrast to the mean field model, because the exact location of bifurcation points is not known for the full system, details such as slopes and scaling exponents can be very much dependent on estimates for the bifurcation point.

\section{Network geometry}

\label{sec:network}

Defining appropriate statistics to capture a fluctuating network
geometry is difficult.  Here, the network does not display
community structure nor is governed by an underlying spatial
structure.  Links are rewired to any acceptable target nodes,
regardless of distance away.  Because the mean field theory for
nodes and links captures the dynamics of the system fairly well,
higher order correlations involving three or more nodes do not
have a large impact on the dynamics.  The network may be fairly
unstructured at the higher levels.

\begin{figure}[tbp]
\begin{center}
\includegraphics[width=4.8in,keepaspectratio]{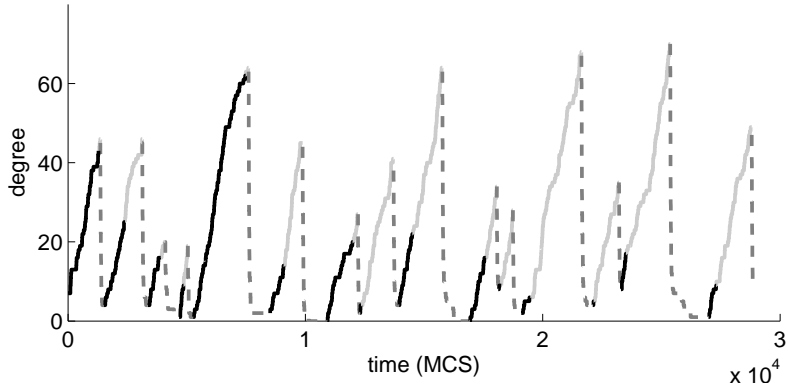}
\end{center}

\caption{Degree of a single node versus time.  Curves indicate the
node's disease status:  black:  recovered; light gray:
susceptible; dashed medium gray:  infected.  Parameters:
 $p=0.002$, $q=0.0016$, $r=0.002$, $w=0.04$.  }
\label{fig:degfluct}
\end{figure}

To demonstrate the role of fluctuations in the network geometry on the scale
of individual nodes and links, we show  in Figure \ref{fig:degfluct} the
time-varying degree of a single arbitrarily chosen node.  When the node
becomes infected, its non-infected neighbors quickly rewire away from it
(dashed gray curves).  Because infected neighbors do not rewire away, the
degree may not drop all the way to zero before pausing.  If the infected node
remains infected for sufficiently long, its neighbors will recover and then
rewire away, further decreasing the degree.  Once the node recovers, other S
and R nodes in the system may rewire to it, and its degree begins to climb
(black curves).  When the node becomes susceptible, the degree continues increasing (solid light gray curves)
until the node again becomes infected, and the cycle repeats. 

\begin{figure}[tbp]
\begin{center}
\includegraphics[width=5in,keepaspectratio]{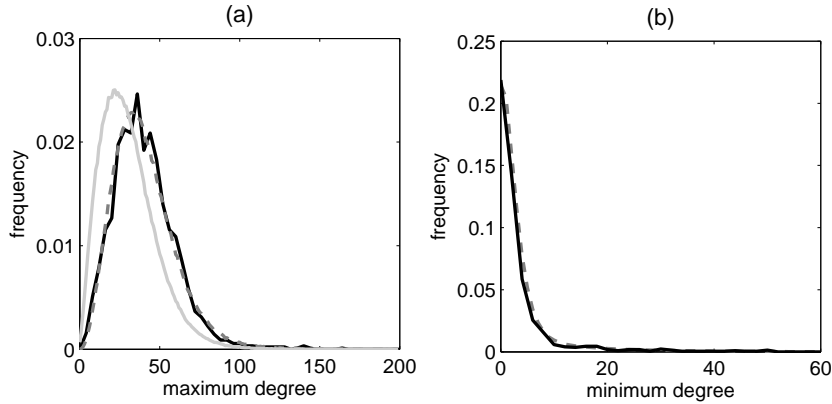}
\end{center}

\caption{Statistics for degree time series. Black curves: observed
distributions; dashed gray curves: expected distributions. (a)
Distribution of local maximum degrees.  The degree distribution
for susceptibles is shown in light gray for reference. (b)
Distribution of local minimum degrees. Parameters:  $p=0.002$,
$q=0.0016$, $r=0.002$, $w=0.04$. } \label{fig:degmaxmin}
\end{figure}

It is not yet known how to predict the degree distributions from
first principles \cite{Shaw08}, but if one assumes that the degree
distributions are already known for each node class, the
fluctuations in the degree of a single node can be easily
understood.  Figure \ref{fig:degmaxmin} shows the statistics of
the local maxima and minima of the degree time series for a single
node. Results are computed for a $4 \times 10^6$ MCS time series,
which contains approximately 2000 SIRS transition cycles (and thus
approximately 2000 maxima and minima).  Frequency distributions
for the maxima (Fig.~\ref{fig:degmaxmin}a) and minima
(Fig.~\ref{fig:degmaxmin}b) are shown.

The minima can be most easily understood.  Minima occur when an
infective recovers.  Recovery is governed by the rate $r$ and is
equally likely to occur for any infective, regardless of the
degree.  Therefore, the distribution for the degree minima is the
same as the degree distribution for infectives.  Figure
\ref{fig:degmaxmin}b shows good agreement between the observed
distribution of minima and that expected from the infective degree
distribution (which was found from Monte Carlo simulations in
\cite{Shaw08}).

Degree maxima occur immediately before a susceptible becomes
infected.  The infection rate depends on the number of infected
neighbors a susceptible node has, which is almost directly
proportional to its degree.  (See \cite{Shaw08} for
details.)  Letting $d$ represent the degree of a susceptible and
$P_d$ the degree distribution for susceptibles, we thus expect the
infection rate to be proportional to $d$ and the distribution of
degree maxima to be proportional to $d P_d$.  This expected
distribution is shown in Figure \ref{fig:degmaxmin}a, and there is
again good agreement with the observed distribution of maxima and
the expected values.  The degree distribution for susceptibles is
shown for reference.  The distribution of maxima is skewed to
higher degrees because of the dependence of the infection process
on node degree.

To develop a complete understanding of these processes, a theory
to predict the degree distributions from first principles is
needed.  Such a theory must account for correlations between the
infection status of an infective and its neighbors, as explained
in \cite{Shaw08}.

\section{Conclusions and discussion}

\label{sec:conclusions}

In this chapter, we considered a model of an adaptive network and its
fluctuations. We introduced a model based on an SIRS epidemic structure
 which included transition probabilities between node states as well as link
 dynamics. In this model, the link dynamics are a function of the
 state variables, and since the state variables depend on the links, it forms a
 closed feedback system between nodes and links. The model is an extension of,
 and contains in the limit of large resusceptibility rate $q$, the SIS model studied in
 \cite{GrossDB06}. The fluctuations  of the  model were simulated in 
 two ways: The full system was studied via Monte Carlo simulation on a
 finite population. In addition, a low dimensional approximation was studied using a Langevin simulation with
 an additive noise term to the mean field equations.

Quantifying where the system is most sensitive to fluctuations required an
examination of the bifurcation structure of the deterministic mean field
equations. For the steady states of the mean field equations, we examined
the locations of both Hopf bifurcations and saddle-node points. We saw that as the resusceptibility rate $q$
changes, the type of bifurcation changes. In general, for large $q$, we have a
generic saddle-node bifurcation, while for small $q$ we have a saddle-saddle
bifurcation giving rise to a Hopf bifurcation. 

Fluctuations were examined with respect to these bifurcation, and particular
attention was paid to the large and small $q$ cases. This led us to examine the
specific role of the recovered class in the SIRS model as compared with the limiting case of the SIS model.  In the SIS
case, we found that without the recovery class, newly created
susceptibles may be still connected to another infective, and thus may become
reinfected immediately. This mechanism led to a reduction in the
fluctuations of infectives. On the other hand, the inclusion of the recovery
class introduced a mean  delay time prior to potential reinfection, thereby
increasing infective fluctuations. 

By examining the fluctuation sizes near the bifurcation points, we found the
existence of scaling laws in both the mean field model and the full system. Comparing
the stochastic dynamics of SIRS and SIS cases, we examined the effect of the
rewiring rate $w$ and resusceptibility rate $q$ on the fluctuations. For a large range
of $q$ values, we showed the existence of a scaling law near the Hopf bifurcation, which
includes the fluctuations for the limiting SIS case. We found that for small
$q$ values, the fluctuations change more rapidly with rewiring rate $w$ than they do for the
SIS model, making fluctuations more sensitive with respect to parameters in the
SIRS case.  Other effects, such as latency, or delay, between infective nodes and
non-infected neighbors occurs in the SIRS model but not the SIS model.

The degree fluctuations are still difficult to predict, although some of
these phenomena can be understood in certain cases. However, much work still
is required to understand the fluctuations of the network geometry. Current
tools for the analysis of static networks are insufficient to make
predictions about adaptive networks.  A complete understanding of the
dynamics, fluctuations, and geometry in the future  requires tools which
incorporate 
topology, stochastic dynamics, and time dependent graph theory. 

Because adaptive networks based on epidemiology  contain many of the features of  adaptive networks in general, we
expect them to continue to reveal new and interesting dynamic phenomena as
they are extended for more detailed modeling of social situations, including
and beyond those of infectious disease spread.

\section{Acknowledgments}
This work was supported by the Office of Naval Research and the Armed Forces Medical
Intelligence Center. LBS was supported by the Jeffress Memorial Trust.

\bibliographystyle{abbrv}
\bibliography{refs}

\begin{thebibliography}{10}

\bibitem{Anderson91}
R.~M. Anderson and R.~M. May.
\newblock {\em Infectious Diseases of Humans.}
\newblock Oxford University Press, 1991.

\bibitem{arnold}
L.~Arnold.
\newblock {\em Random Dynamical Systems}.
\newblock Springer, New York, 2001.

\bibitem{Barabasi99}
A.~Barab\'{a}si and R.~Albert.
\newblock Emergence of scaling in random networks.
\newblock {\em Science}, 286(5439):509--512, 1999.

\bibitem{Benczik07}
I.~J. Benczik, S.~Z. Benczik, B.~Schmittmann, and R.~K.~P. Zia.
\newblock Lack of consensus in social systems.
\newblock http://arxiv.org/abs/0709.4042, 2007.

\bibitem{Bornholdt03}
S.~Bornholdt and T.~R\"{o}hl.
\newblock Self-organized critical neural networks.
\newblock {\em Physical Review E}, 67(6):066118, 2003.

\bibitem{Bornholdt98}
S.~Bornholdt and K.~Sneppen.
\newblock Neutral mutations and punctuated equilibrium in evolving genetic
  networks.
\newblock {\em Physical Review Letters}, 81(1):236--239, 1998.

\bibitem{Christensen98}
K.~Christensen, R.~Donangelo, B.~Koiller, and K.~Sneppen.
\newblock Evolution of random networks.
\newblock {\em Physical Review Letters}, 81(11):2380, 1998.

\bibitem{CostaRTB07}
L.~D. Costa, F.~A. Rodrigues, G.~Travieso, and P.~R.~V. Boas.
\newblock Characterization of complex networks: A survey of measurements.
\newblock {\em Advances In Physics}, 56:167--242, 2007.

\bibitem{auto}
E.~J. Doedel, R.~Paffenroth, A.~Champnets, T.~Fairgrieve, Y.~A. Kuznetsov,
  B.~Sandstede, and X.~Wang.
\newblock {\em AUTO: Software for continuation and bifurcation for ordinary
  differential equations}, 2001.

\bibitem{Dykman1980}
M.~I. Dykman and M.~A. Krivoglaz.
\newblock Fluctuations in non-linear systems near bifurcations corresponding to
  the appearance of new stable states.
\newblock {\em Physica A}, 104(3):480--494, 1980.

\bibitem{Ebel02}
H.~Ebel and S.~Bornholdt.
\newblock Coevolutionary games on networks.
\newblock {\em Phys. Rev. E}, 66(5):056118, 2002.

\bibitem{Ehrhardt06}
G.~C. M.~A. Ehrhardt, M.~Marsili, and F.~V. Redondo.
\newblock Phenomenological models of socioeconomic network dynamics.
\newblock {\em Physical Review E (Statistical, Nonlinear, and Soft Matter
  Physics)}, 74(3):036106, 2006.

\bibitem{EhrhardtMV06}
G.~C. M.~A. Ehrhardt, M.~Marsili, and F.~Vega-Redondo.
\newblock Phenomenological models of socioeconomic network dynamics.
\newblock {\em Physical Review E}, 74:036106, 2006.

\bibitem{Fan04}
Z.~Fan and G.~Chen.
\newblock Evolving networks driven by node dynamics.
\newblock {\em Internationl. Journal of Modern Physics B}, 18:2540--2546, 2004.

\bibitem{Gil06}
S.~Gil and D.~H. Zanette.
\newblock Coevolution of agents and networks: Opinion spreading and community
  disconnection.
\newblock {\em Physics Letters A}, 356(2):89--94, 2006.

\bibitem{Gleiser06}
P.~M. Gleiser and D.~H. Zanette.
\newblock Synchronization and structure in an adaptive oscillator network.
\newblock {\em European Physics Journal B}, 53:233--238, 2006.

\bibitem{Gong04}
P.~Gong and C.~van Leeuwen.
\newblock Evolution to a small-world network with chaotic units.
\newblock {\em Europhys. Lett.}, 67:328--333, 2004.

\bibitem{Graham1987a}
R.~Graham and T.~T\'el.
\newblock Nonequilibrium potentials for local codimension-2 bifurcations of
  dissipative flows.
\newblock {\em Physical Review A}, 35(3):1328--1349, 1987.

\bibitem{Gross07}
T.~Gross and B.~Blasius.
\newblock Adaptive coevolutionary networks: a review.
\newblock {\em Journal of the Royal Society. Interface}, 2007.
\newblock DOI: 10.1098/rsif.2007.1229.

\bibitem{GrossDB06}
T.~Gross, C.~J.~D. D'Lima, and B.~Blasius.
\newblock Epidemic dynamics on an adaptive network.
\newblock {\em Physical Review Letters}, 96:208701, 2006.

\bibitem{Holme06}
P.~Holme and G.~Ghoshal.
\newblock Dynamics of networking agents competing for high centrality and low
  degree.
\newblock {\em Physical Review Letters}, 96(9):098701, 2006.

\bibitem{HolmeN06}
P.~Holme and M.~E.~J. Newman.
\newblock Nonequilibrium phase transition in the coevolution of networks and
  opinions.
\newblock {\em Physical Review E}, 74(5):056108, 2006.

\bibitem{Horsthemke83}
W.~Horsthemke and R.~Lefever.
\newblock {\em Noise-Induced Transitions: Theory and Applications in Physics,
  Chemistry, and Biology}.
\newblock Springer Series in Synergetics , Vol. 15, 1983.

\bibitem{Ito02}
J.~Ito and K.~Kaneko.
\newblock Spontaneous structure formation in a network of chaotic units with
  variable connection strengths.
\newblock {\em Physical Review Letters}, 88(2):028701, 2002.

\bibitem{Ito03}
J.~Ito and K.~Kaneko.
\newblock Spontaneous structure formation in a network of dynamic elements.
\newblock {\em Physical Review E}, 67(4):046226, 2003.

\bibitem{Jain01}
S.~Jain and S.~Krishna.
\newblock A model for the emergence of cooperation, interdependence, and
  structure in evolving networks.
\newblock {\em Proceedings of the National Academy of Science}, 98:543--547,
  2001.

\bibitem{Newman03}
M.~E.~J. Newman.
\newblock The structure and function of complex networks.
\newblock {\em SIAM Review}, 45(2):167--256, 2003.

\bibitem{PachecoTN06}
J.~M. Pacheco, A.~Traulsen, and M.~A. Nowak.
\newblock Coevolution of strategy and structure in complex networks with
  dynamical linking.
\newblock {\em Physical Review Letters}, 97:258103, 2006.

\bibitem{Scholz07}
J.~C. Scholz and M.~O.~W. Greiner.
\newblock Topology control with ipd network creation games.
\newblock {\em New Journal of Physics}, 8:185--199, 2007.

\bibitem{Shaw08}
L.~B. Shaw and I.~B. Schwartz.
\newblock Fluctuating epidemics on adaptive networks.
\newblock {\em Physical Review E}, 77:066101, 2008.

\bibitem{SkyrmsP00}
B.~Skyrms and R.~Pemantle.
\newblock A dynamic model of social network formation.
\newblock {\em Proceedings of the National Academy of Sciences}, 97:9340--9346,
  2000.

\bibitem{Vazquez08}
F.~Vazquez, V.~M. Egu\'{\i}luz, and M.~San~Miguel.
\newblock Generic absorbing transition in coevolution dynamics.
\newblock {\em Physical Review Letters}, 100(10):108702, 2008.

\bibitem{ZanetteG06}
D.~H. Zanette and S.~Gil.
\newblock Opinion spreading and agent segregation on evolving networks.
\newblock {\em Physica D-Nonlinear Phenomena}, 224:156--165, 2006.

\bibitem{Zanette07}
D.~H. Zanette and S.~R. Gusman.
\newblock Infection spreading in a population with evolving contacts.
\newblock http://arxiv.org/abs/0711.0874, 2007.

\bibitem{Zhou06}
C.~Zhou and J.~Kurths.
\newblock Dynamical weights and enhanced synchronization in adaptive complex
  networks.
\newblock {\em Physical Review Letters}, 96(16):164102, 2006.

\end{thebibliography}

\end{document}